\documentclass[]{article}
\usepackage{times}
\usepackage{color}
\usepackage{graphicx}
\usepackage{graphics}
\usepackage{amsmath} 
\usepackage{amssymb}  
\usepackage{amsfonts} 
\usepackage[dvips]{epsfig}

\begin{document}

\vspace*{1.7cm}

{\Large {\bf Particle-scale origins of shear strength in granular media}}\\

{\large Farhang Radjai}\\

LMGC, CNRS-Universit\'e Montpellier 2, CC048, 34095 Montpellier, France.\\
radjai@lmgc.univ-montp2.fr. 

\vspace*{0.5cm}

\rule{12.5cm}{0.5pt}\\
{\small {\bf Abstract}.
The shear strength of cohesionless granular materials is generally attributed to the compactness or anisotropy of their microstructure. 
An open issue is how such compact or anisotropic 
microstructures, and thus the shear strength, depend on the particle properties. 
We first recall the role of fabric and force anisotropies with  
respect to the critical-state shear stress.  
Then, a model of accessible geometrical states in terms of particle connectivity and contact anisotropy is presented. This model incorporates in a 
simple way the fact that, due to steric exclusions, the  highest levels of connectivity and 
anisotropy cannot be reached simultaneously, a property that affects seriously the shear strength. We also analyze the force anisotropy in the light of 
the specific role of weak forces in sustaining strong force chains and thus the main mechanism that underlies anisotropic force patterns. Finally, we briefly 
discuss the effect of interparticle friction, 
particle shape, size polydispersity and adhesion.  
 \\ 
 
{\bf keyword:} granular media; shear strength; fabric anisotropy; 
weak and strong forces.

}

\rule{12.5cm}{0.5pt}


\section{Introduction}

Since the early work of Coulomb in 1773, 
the plastic yield behavior of granular materials has remained 
an active research field in close connection  
with soil mechanics and powder technology \cite{Mitchell2005, Nedderman1992}. 
According to the Mohr-Coulomb yield criterion, for normal and shear stresses 
$\sigma$ and $\tau$ acting on a slip plane, the plastic threshold $\tau_c$ is 
the sum of two terms: 
\begin{equation}
\tau_c = c+\sigma \tan \varphi,
\label{eq:tau}
\end{equation}
where $c$ is a cohesive strength and $\varphi$ is the internal angle of 
friction depending only on the nature of the granular material. 
This criterion expresses the pressure dependence of shear 
strength which is a distinctive feature of granular media. Given (\ref{eq:tau}), 
the shear strength of cohesionless materials 
($c=0$) can be represented by the (dimensionless) stress ratio 
$\tau_c / \sigma = \mu_c = \tan\varphi$. Since the  angle $\varphi$ is 
a bulk property, it can be expressed in terms of stress invariants. Let  $\sigma_\alpha$ ($\alpha = 1,2,3$) be stress principal values. 
The average stress is $p= (\sigma_1 + \sigma_2)/2$ in 2D and 
$p= (\sigma_1 + \sigma_2 + \sigma_3)/3$ in 3D. We define the stress deviator by 
$q=(\sigma_1 - \sigma_2)/2$ in 2D and $q= (\sigma_1 - \sigma_3)/3$ in 3D under axisymmetric 
conditions ($\sigma_2=\sigma_3$). With these notations, it can be shown that  
$\sin \varphi = q/p$ in 2D and $\sin \varphi = 3q/(2p+q)$ in 3D. 

\begin{figure}[!bp]
\centerline{\includegraphics[width=8cm]{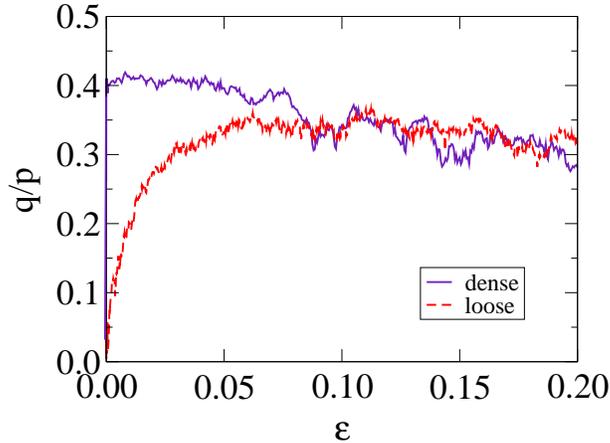}}
\caption{Normalized shear stress as a function of cumulative shear strain in 
a 2D simple shear simulation by the contact dynamics 
method for two different values of the initial solid fraction. 
\label{fig:qp}}
\end{figure}

This picture of  shear strength in granular media   
holds as a basic fact although  
the complex plastic behavior of granular media can 
not be reduced to a single strength parameter. In particular, 
the shear strength and plastic flow (dilatancy) depend on 
the granular structure and direction of loading, the latter reflecting  
the anisotropy of the structure.  
Since the shear strength is state-dependent,  it cannot be considered as a 
material property unless attributed to a well-defined granular state. 
The internal angle of friction $\varphi$ is often associated with  
the critical state (steady state or residual state) reached after long monotonous 
shearing; see Fig. \ref{fig:qp}. This state is characterized by  
a solid fraction $\rho_c$ independent of the loading history and 
initial conditions \cite{Wood1990}. 

The critical-state strength is below the  peak shear stresses
occurring for dense states with solid fraction $\rho_0 > \rho_c$, 
but these states are metastable and often lead to 
strain localization \cite{Darve2000, Vardoulakis1995}.  
For loose states with $\rho_0 < \rho_c$, the 
critical state is reached asymptotically  following 
diffuse rearrangements. Hence, apart from these  
transients, which are governed by the evolution of 
internal variables pertaining to the microstructure and are important 
in formulating elasto-plastic models, the critical-state shear strength 
represents a stable plastic threshold for a granular material.    
    
In this paper, we are interested in the critical-state strength as a 
material property of cohesionless granular materials. The critical-state  
friction angle $\varphi_c$  can be described 
as a coarse-grained (or homogenized) friction angle between 
two granular layers sliding past each other. Nevertheless, the macroscopic 
status of $\varphi_c$ as 
a Coulomb friction angle, on the same grounds as those of dry 
friction between solid bodies, should not eclipse the fact that 
the granular friction angle is a 
bulk property to which adequate tensorial stress analysis should be applied 
(this was indeed the contribution of Mohr) and where the slip planes are 
not {\em a priori} defined,  in contrast to solid friction which is 
a surface property at the macroscopic scale \cite{Radjai2004a}. Depending on the boundary conditions, 
the critical state occurs either homogeneously 
in the whole volume of a granular sample or inside a thick layer of several 
particle diameters in the advent of strain 
localization \cite{Bardet1992, Herrmann1995c, Vermeer1990, Moreau1997a}. In both configurations, 
$\varphi_c$ stems from various granular 
phenomena such as friction between particles, anisotropy of the microstructure, 
organization of force networks and dissipation due to inelastic collisions. 
We consider below these effects and their respective roles 
in enhancing or restraining granular friction.


\section{Effect of interparticle friction}
\label{sec:friction}

While solid friction between particles underlies the frictional 
behavior of granular materials, it is not obvious how and through which physical mechanisms it comes into play. If shear deformation took place as a result 
of sliding  between all contacts  along a slip plane, the 
friction angle $\varphi_c$ would simply echo the friction between particles. 
An example of such a configuration is a regular pile of cubic blocs subjected to 
a vertical load. Horizontal shearing of this pile implies sliding between at least 
two rows so that  the shear strength of the pile is a straightforward effect of  
friction between the blocs. However, discrete numerical simulations suggest that 
in sheared granular materials, rolling prevails over sliding \cite{Radjai1998}. 
In quasistatic shear, 
sliding occurs at only $\simeq 10\%$ of contacts, and  these sliding contacts belong 
essentially to weak contacts (see below) oriented on average along 
the minor principal stress directions \cite{Radjai1999, Staron2005, Staron2005a}. Hence, the relationship between $\varphi_c$ and 
the local friction angle $\varphi_s$ involves the inhomogeneous distribution of forces and mobilization (or activation) of the friction force at rolling contacts.                        

\begin{figure}[!bp]
\centerline{\includegraphics[width=8cm]{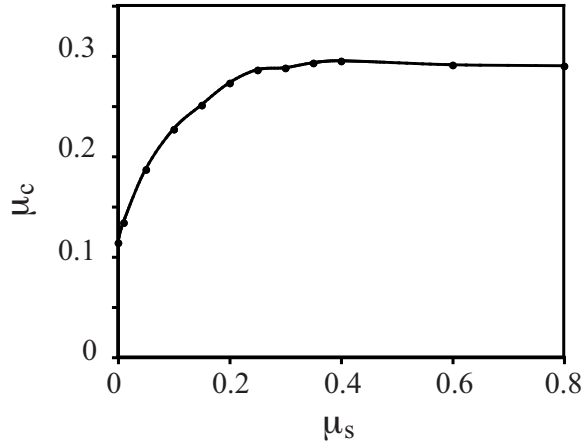}}
\caption{The critical-state friction coefficient $\mu_c$ as a 
function of sliding friction coefficient $\mu_s$ between particles in biaxial 
shearing of a sample of 5000 particles. 
\label{fig:mus_muc}}
\end{figure}

This relationship is far from linear as shown in Fig. \ref{fig:mus_muc}.  
The critical-state coefficient $\mu_c = \tan \varphi_c$ is 
above $\mu_s = \tan \varphi_s$ at small values of the latter, 
and at larger values it tends to a plateau $ \mu_\infty < \mu_s$ \cite{Corriveau1997, Taboada2006}. 
The transition from 
$\mu_c - \mu_s <0$ to $\mu_c - \mu_s >0$ occurs 
at $\mu_c = \mu_s \simeq 0.5$. Beyond $\mu_s = 0.5$,  
$\mu_c$ is practically independent of $\mu_s$. 
The independence 
of $\varphi_c$ with respect to $\varphi_s$ at 
large values of the latter indicates that the role of interparticle  friction 
is more subtle than expected from simple models. 
Moreover, the nonzero value of $\varphi_0$ shows clearly that 
the interparticle friction is not the only source of 
frictional behavior in the critical state \cite{Roux2001}.  

The {\em direct} contribution of interparticle friction to 
shear strength, i.e. without interposition by the microstructure as 
will be analyzed below,  may be evaluated from a decomposition 
of the shear stress. The stress tensor $\sigma_{\alpha\beta}$ in a control 
volume $V$ can be expressed 
as \cite{Rothenburg1981, Christoffersen1981, Moreau1997a,
 Bagi1999, Staron2005a}
\begin{equation}
\centering
\sigma_{\alpha\beta}= n_b \langle \ell_\alpha^i f_\beta^i \rangle, 
\label{eq:sig}
\end{equation}
where $n_b$ is the number density of bonds (contacts), $\ell_\alpha^i$ is the 
$\alpha$-component of the branch vector ${\boldsymbol \ell}^i$  
joining the centers of particles at contact $i$ and  $f_\beta^i$ 
is the $\beta$-component of the 
force vector $\boldsymbol f$ acting at the contact $i$ between the two particles. 

The contribution of friction forces can be estimated by replacing 
 in equation (\ref{eq:sig}) 
$\boldsymbol f$ by ${\boldsymbol f} \cdot {\boldsymbol t} \ {\boldsymbol t}$,  
where $\boldsymbol t$ is the unit vector along the friction 
force. The contribution of normal forces is the complementary tensor obtained by 
replacing  $\boldsymbol f$ by ${\boldsymbol f} \cdot {\boldsymbol n} \ {\boldsymbol n}$,  where $\boldsymbol n$ is the 
unit vector perpendicular to the contact plane. The corresponding shear strengths 
$q_t$ and $q_n$ can then be calculated in the critical state. 
Numerical simulations show that the ratio $q_t / q$ is 
quite low (below $10\%$)\cite{Cambou1993}. 
This counterintuitive finding underlines the role of interparticle friction  
as a parameter acting ``behind the scenes" rather than a direct  
actor of shear strength. Our simulations show that,  due to 
disorder and force/moment balance conditions as well as kinematic constraints such 
as rotation frustration,  
the friction forces inside a granular packing are strongly coupled with 
normal forces. For example, highly mobilized friction forces are rare events and 
the distribution of friction forces reflects for the most part 
that of normal forces. We consider below such effects in 
connection with granular microstructure.         


\section{Harmonic representation of the microstructure} 
The microscopic expression of the stress tensor in equation (\ref{eq:sig}) 
is an arithmetic mean involving  
the branch vectors and contact forces. Hence,  
for analyzing the particle-scale origins of the shear strength, 
we need a statistical description of the granular microstructure and 
force transmission. Noticing that the shear stress corresponds to
the deviation of stress components from the mean stress 
$p=tr (\boldsymbol \sigma)/d$ (for space dimension $d$) 
along different space directions, the useful 
information for this analysis is the density and average force 
of all contacts pointing in the same direction as a function of 
this direction. These functions can be expanded in Fourier series in 2D and in spherical 
harmonics in 3D\cite{Rothenburg1989, Ouadfel2001}. 
Since the contacts have no polarity, the period is $\pi$. 

For illustration, we consider here only the 2D expansions 
truncated beyond the second term:
\begin{equation}
\left\{
\begin{array}{lcl}
P_\theta (\theta) &=& \frac{1}{\pi}  \{ 1 + a \cos 2(\theta - \theta_b) \},  \\
\langle f_n \rangle (\theta) &=& \langle f \rangle  \{ 1 + a_n \cos 2(\theta - \theta_n) \},  \\
\langle f_t \rangle (\theta) &=&  \langle f \rangle  a_t \sin 2(\theta - \theta_t) , 
\end{array}
\right.
\label{eq:Pff}
\end{equation}  
where  $P_\theta$ is the probability density function of contact normals, and $f_n$ and $f_t$ 
are the force components along (radial) and perpendicular 
to (orthoradial) the branch vector, respectively. 
The parameters $a$, $a_n$ and $a_t$ are the anisotropies of 
branch vectors, radial forces and orthoradial forces, respectively, 
$\theta_b$, $\theta_n$ and $\theta_t$ being the corresponding privileged 
directions.  The sine function for the expansion of the orthoradial 
component $f_t$  is imposed by the requirement that 
the mean orthoradial force is zero to satisfy the balance of force moments over particles whereas the mean radial force  $\langle f \rangle$ is positive (repulsive). We also 
note that for circular and spherical particles the radial and orthoradial force 
components coincide with normal and tangential forces, respectively.  

This {\em harmonic representation} with only three anisotropy parameters provides 
a good approximation for numerical data. 
Using the functions (\ref{eq:Pff}), the stress components $\sigma_{\alpha\beta}$ can 
be written as an integral over space directions:  
\begin{equation}
\sigma_{\alpha \beta} = 
n_b \langle \ell \rangle \int_0^\pi  \left\{ \langle f_n \rangle (\theta) n_\alpha(\theta) + 
\langle f_t \rangle (\theta) t_\beta(\theta) \right \} P_\theta(\theta) \ d\theta,   
\label{eq:sigint}
\end{equation}
where  $n_x = \cos(\theta)$ and $n_y = \sin(\theta)$, $t_x = -\sin(\theta)$ and 
$t_y = \cos(\theta)$. It has been also assumed that the branch vector 
lengths $\ell$ are not correlated with forces.   

Equation (\ref{eq:sigint}) together with the harmonic approximation  
expressed in equation (\ref{eq:Pff}) yield the following expression for the 
normalized stress deviator \cite{Radjai2004a}: 
\begin{equation}     
\frac{q}{p} \simeq \frac{1}{2} \left\{ a \cos 2(\theta_\sigma - \theta_b) 
+a_n \cos 2(\theta_\sigma - \theta_n) +a_t\cos 2(\theta_\sigma - \theta_t) \right\},
\label{eq:qa}
\end{equation}
where $\theta_\sigma$ is the major principal direction of the stress tensor. 
In deriving equation (\ref{eq:qa}), the cross products among the anisotropies 
have been neglected.   
In the critical state, the privileged directions 
coincide, i.e. $\theta_b \simeq \theta_n \simeq \theta_t \simeq \theta_\sigma$, 
so that \cite{Rothenburg1989, Ouadfel2001} 
\begin{equation}     
\frac{q_c}{p} \simeq \frac{1}{2} \left\{ a_c + a_{nc} +a_{tc} \right\},  
\label{eq:qcp}
\end{equation}
where the anisotropy parameters refer to the critical state. In 3D, a similar relation can be established by means of spherical harmonics \cite{Azema2008}:
\begin{equation}     
\frac{q_c}{p} \simeq \frac{2}{5} \left\{ a_c + a_{nc} +a_{tc} \right\} 
\label{eq:qcp3D}
\end{equation}

These relations  exhibit  two 
microscopic sources of the shear strength in a granular packing: 
1) fabric anisotropy, represented by the parameter $a$ and 
2) force anisotropy,  captured into the parameters $a_n$  and $a_t$. 
Hence,  the material parameters influence  
the shear strength via fabric and force anisotropies. For example, 
the saturation of $\varphi_c$  for $\varphi_s > 0.5$ 
(section \ref{sec:friction}
means that, increasing the interparticle friction beyond this limit 
does not enhance anisotropy. 
    
\section{Accessible geometrical states}
\label{sec:acc}

In this section, we focus on the fabric anisotropy $a$ which represents 
the excess and loss of contacts along  different space directions with 
respect to the average contact density. The latter is commonly represented 
by the coordination number $z$ (mean number of contacts per particle). 
In a granular material, $z$ is bounded between two limits $z_{min}$ and    
$z_{max}$. The lower bound $z_{min}$ is dictated by the force balance 
requirement. For example, stable particles often involve 
more than three contacts in 2D and more than four contacts in 3D. 
On the other hand, the upper bound  $z_{max}$ is constrained by steric 
exclusions \cite{Troadec2002}. 
For example, in 2D for a system of monodisperse particles, a particle 
can not have more than 6 contacts. In practice, this limit is reduced to 4 
as a result of disorder. 

Within the harmonic approximation, the geometrical state of a granular system 
is defined by its position in the space of coordinates $(z,a)$. We define two 
limit states: 1) the loosest isotropic state, characterized by $(z=z_{min}, a=0)$, and 
2) the densest isotropic state, characterized by   $(z=z_{max}, a=0)$. 
These states can be reached only by complex loading. For example,     
it is generally difficult to bring a granular system towards a 
dense isotropic state via isotropic compaction.  
The reason is that the rearrangements occur mainly in the presence of shearing, and the latter leads to fabric anisotropy. 

\begin{figure}[!bp]
\centerline{\includegraphics[width=8cm]{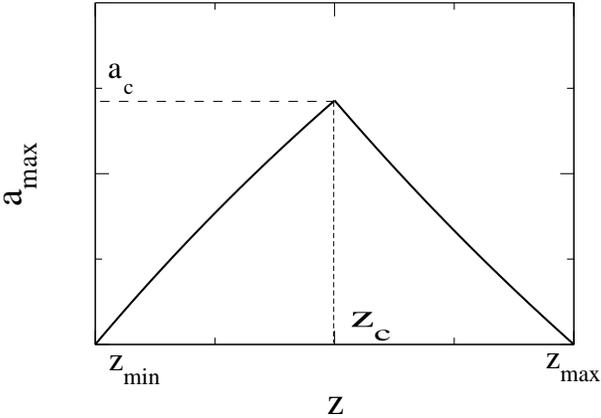}}
\caption{Domain of accessible geometrical states 
based on the harmonic representation of granular microstructure.
\label{fig:amax}}
\end{figure}

It is natural to assume that all accessible geometrical states are enclosed between 
the two isotropic  limit states. In order to represent the geometrical 
states, it is useful to define the {\em state function} 
\begin{equation}
E(\theta)= z P_\theta (\theta) = \frac{z}{ \pi } \{ 1 + a \cos 2 (\theta - \theta_b)  \}.
\label{eq:E}
\end{equation}
The two limit isotropic states are $E_{min} = z_{min} / \pi$ and
$E_{max} = z_{max} / \pi $. The assumption that the geometrical states are constrained 
to stay between the two isotropic limit states, implies that  the 
anisotropy $a$ can not exceed a maximum $a_{max}$ depending on the 
value of $z$. With harmonic approximation (\ref{eq:E}), we obtain
\begin{equation}
a_{max}(z) = \mbox{min} \left\{ 2 \left(  1-\frac{z_{min}}{z}  \right) , 2 \left(  \frac{z_{max}}{z} - 1  \right)    \right\}.
\label{eq:amax}
\end{equation}
This function is shown in Fig. \ref{fig:amax}.  By construction, 
$a_{max}(z_{min}) = a_{max}(z_{max}) = 0$. 
The largest anisotropy  is
\begin{equation} 
a_{c} =  a_{max} (z_{c}) = 2 \frac{ a_{max} - a_{min} } {a_{max}+a_{min} },
\label{eq:ac}
\end{equation}
with $z_{c} = (z_{min}+z_{max})/2$.  
According to equation (\ref{eq:ac}),  $a_{max}$ increases 
with $z$  for $z < z_{c}$, and it declines with $z$   for  $z < z_{c}$.  
When $a=a_{c}$  is reached along a monotonic path, 
neither anisotropy nor coordination number evolve 
since both contact gain and contact loss are saturated.  
In this picture, the critical state corresponds to the intersection  
between the two regimes with $z=z_{c}$ and $a=a_{c}$. 
In 2D with weakly polydisperse circular particles and $\mu_s > 0.5$, 
a good fit is 
provided by assuming $z_{min}=3$ and $z_{max}=4$. This yields $z_c= 3.5$ and 
$a_c=2/7$. For lower values of $\mu_s $, $a_c$ declines. 

\begin{figure}[!bp]
\centerline{\includegraphics[width=8cm]{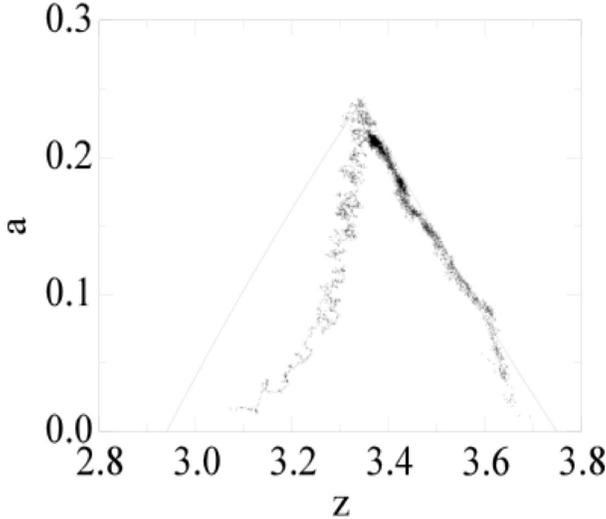}}
\caption{Evolution of the geometrical state of a sheared  
packing for two different initial states simulated by the contact dynamics method. 
\label{fig:az}}
\end{figure}

Fig. \ref{fig:az} shows the evolution of $a$ with $z$ in simulated biaxial compression 
of two initially isotropic samples 
with initial coordination numbers $z_0 = 3.1$ and $z_0 = 3.7$. 
In both simulations, 
$z$ tends to the same critical-state value 
$z_c \simeq 3.35$ with $a_c \simeq 0.24$. Remarkably,  
the anisotropy of the dense packing reaches and 
then follows closely the limit states. 
Equation (\ref{eq:amax}) provides here an excellent fit to the data 
with only one fitting parameter $z_{max}$. In the loose case, 
the trajectory remains entirely inside the domain of accessible states and the limit 
states are reached only at the critical state  

Equation (\ref{eq:ac}) predicts that the critical state anisotropy $a_c$ 
increases with $z_{max} - z_{min}$. The shape, size and frictional characteristics 
of the particles may therefore influence $a_c$ via  $z_{min}$ and $z_{max}$. 
For example, increasing the sliding friction between the particles 
allows for lower values of  
$z_{min}$ (stable configurations with less contacts) without 
changing $z_{max}$ (which depends only on steric exclusions) 
and leads to larger values of $a_c$. 

One interesting aspect of the model of accessible states presented in 
this section is  to show that the largest values of $a$ and $z$ can not 
be reached simultaneously. The critical value $a_c$ is not obtained 
with $z_{max}$ but with $z_c$ which is below $z_{max}$. But  
higher levels of force anisotropies $a_{nc}$ and $a_{tc}$ can be achieved 
with higher values of $z$.        

\begin{figure}[!bp]
\centerline{\includegraphics[width=6cm,angle=-90]{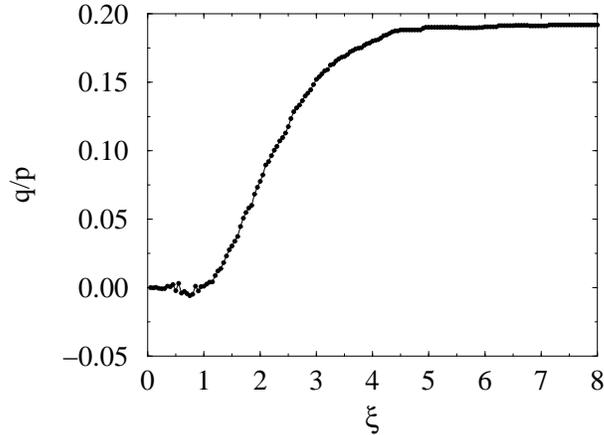}}
\caption{The partial shear stress, normalized by the mean stress, as a 
function of force threshold $\xi$. 
\label{fig:qxi}}
\end{figure}

\begin{figure}[!bp]
\centerline{\includegraphics[width=8cm,angle=-90]{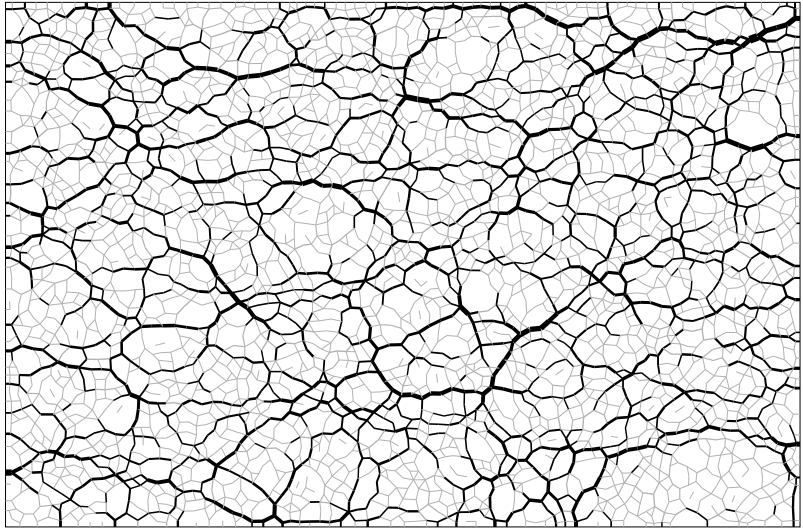}}
\caption{Weak and strong normal forces represented in 
two different grey levels. Line thickness is proportional to 
the normal force. 
\label{fig:forces}}
\end{figure}


\section{Weak and strong force networks}

According to equation (\ref{eq:ac}), 
the shear strength is proportional to force anisotropies $a_{nc}$ and $a_{tc}$ 
in the critical state. As for fabric anisotropy $a$, which was discussed in the 
last section, we would like to analyze here  the mechanisms that underly 
force anisotropies.   
A basic feature of force distribution in granular media 
is the occurrence of numerous weak forces together with a subnetwork 
of strong forces appearing often sequentially (force chains). 
The probability density function (pdf) $P_n(f_n)$   
of normal forces in a macroscopically homogeneous system in the critical state 
is such  that 
more than $58\%$  of contact forces are below the mean force  $\langle f_n \rangle$  
and they have a nearly uniform distribution 
\cite{Radjai1996a, Mueth1998, Tsoungui1998}.  These {\em weak} forces 
contribute only $\simeq 29\%$  to the mean stress $p$. 
The pdf of {\em strong} forces (above the mean normal force $\langle f_n \rangle$) 
decays exponentially {Radjai1996a, Coppersmith1996a, Radjai1999, Majmudar2005, Metzger2004}.  
The very large number of weak forces, reflecting the arching effect, 
is a source of weakness for the system.  
Weak regions inside a packing correspond to locally weak pressures and they   
are more susceptible to fail.  A quantitative analysis of grain rearrangements 
indicates that during a quasistatic evolution those weak regions undergo local rearrangements, and nearly all sliding contacts  are localized in weak 
regions \cite{Staron2002, Staron2005a, Nicot2006}.

Let ${\cal S}(\xi)$ be the set of contacts with a normal 
force $f_n <  \xi \langle f_n \rangle$.  
The set ${\cal S}(\infty)$  is the whole contact set. The weak and strong sets 
are ${\cal S}(1)$ and ${\cal S}(\infty) - {\cal S}(1)$, respectively. 
The partial shear stress $q(\xi) /p$ and the 
fabric and force anisotropies $a(\xi)$, $a_n(\xi)$ and $a_t(\xi)$ 
can be calculated as a function of $\xi$ \cite{Radjai1998}. 
Our simulations show that $q(\xi) \simeq 0$; see Fig. \ref{fig:qxi}.  
This means that  nearly the whole stress deviator is carried by 
the strong contact network, the weak contacts contributing only to 
the mean stress.  Hence, the total stress tensor $\boldsymbol \sigma$ is a
sum of two terms:
\begin{equation}
{\boldsymbol \sigma} = p_w {\boldsymbol I} + {\boldsymbol \sigma}_s,
\label{sigw}
\end{equation}
where ${\boldsymbol I}$ is the unit tensor, $p_w$ is the weak pressure, and 
${\boldsymbol \sigma}_s$ represents the strong stress tensor. 
Hence, from the  stress transmission viewpoint, 
the weak contact set is a ``liquidlike" phase whereas 
the strong contact set appears as a ``solidlike" backbone 
transmitting shear stresses. The weak and strong networks are shown in 
Fig. \ref{fig:forces}  in thickness 
of segments joining particle centers for an assembly of 4000 particles 
subjected to biaxial compression.

The zero shear stress in the weak network implies that, according to 
equation (\ref{eq:qcp}), at least one of the corresponding 
anisotropies is negative. Since 
the critical-state angles are assumed to be equal 
($\theta_b \simeq \theta_n \simeq \theta_t \simeq \theta_\sigma$), a negative 
value corresponds to a rotation $\pi /2$ of the principal axes.   
Indeed, our numerical data show that the privileged direction of weak contacts 
is perpendicular to the major principal stress direction \cite{Radjai1998}. 
The strong forces occur at contacts that are, on average, aligned with the major 
principal 
direction of the stress tensor. Lateral weak forces prop the particles 
against deviations from alignment at strong contacts. In other words, 
the weak contacts play the same stabilizing role with respect to the 
particles sustaining strong forces as the counterforts 
with respect to an architectural arch. This {\em bimodal} transmission 
of shear stresses corresponds thus to a statistical description of  
arching effect in granular media.      

This stress-fabric correlation can be interpreted as a 
way for a granular system to optimize the shear strength. 
Indeed, the stress deviator $q$ increases if a larger number of strong forces 
occur at contacts aligned with the major principal direction, implying thus 
a surplus of weak contacts in the perpendicular direction. This {\em weakening} 
of forces at contacts pointing in one direction has the same effect for  
force anisotropy as the loss of contacts in the same direction 
for fabric anisotropy. As a result, force weakening in the weak 
network  is all the more efficient as it leads to lower amount of contact loss. 
This condition can, for example, be achieved for higher level 
of connectivity, i.e. larger values of $z$ in the critical state.  

\section{Effect of material parameters}
\label{sec:eff}

In this section, we briefly discuss the effect of several material 
parameters with respect to the mechanisms that underly shear 
strength in granular media. More details will be given elsewhere. 

There are several shape parameters that may lead to enhanced shear strength 
through force anisotropy or fabric anisotropy. We consider here polygonal 
particles as compared to circular particles \cite{Azema2007}. The first sample, denoted  
S1, is composed of 14400 regular pentagons of three different diameters: 
$50\%$ of diameter $2.5$ cm, $34\%$ of diameter $3.75$ cm and $16\%$ of 
diameter $5$ cm. The second sample, denoted S2, is composed of 10000 disks 
with the same polydispersity. 
The coefficient of friction is 0.4 between particles and 0 with the walls. 
At equilibrium, both numerical samples are in  isotropic stress state.  
The solid fraction is $0.80$ for S1 and  $0.82$ for S2. 
The isotropic samples are subjected to vertical compression by downward 
displacement of the top wall.

\begin{figure}
\centerline{\includegraphics[width=8cm]{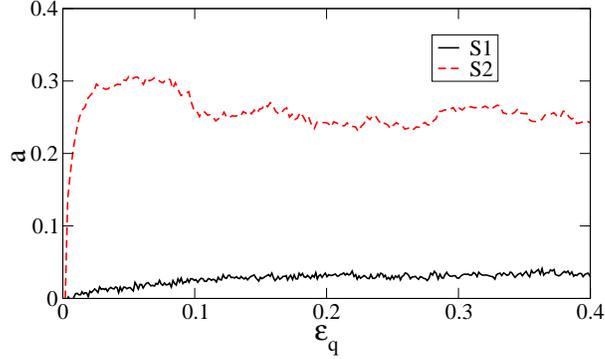}}
\caption{Evolution of the anisotropy $a$ with  cumulative shear strain $\varepsilon_q$
for a packing of pentagons (S1) and a packing of disks (S2).    \label{fig:ae}}
\end{figure}

\begin{figure}
\centerline{\includegraphics[width=8cm]{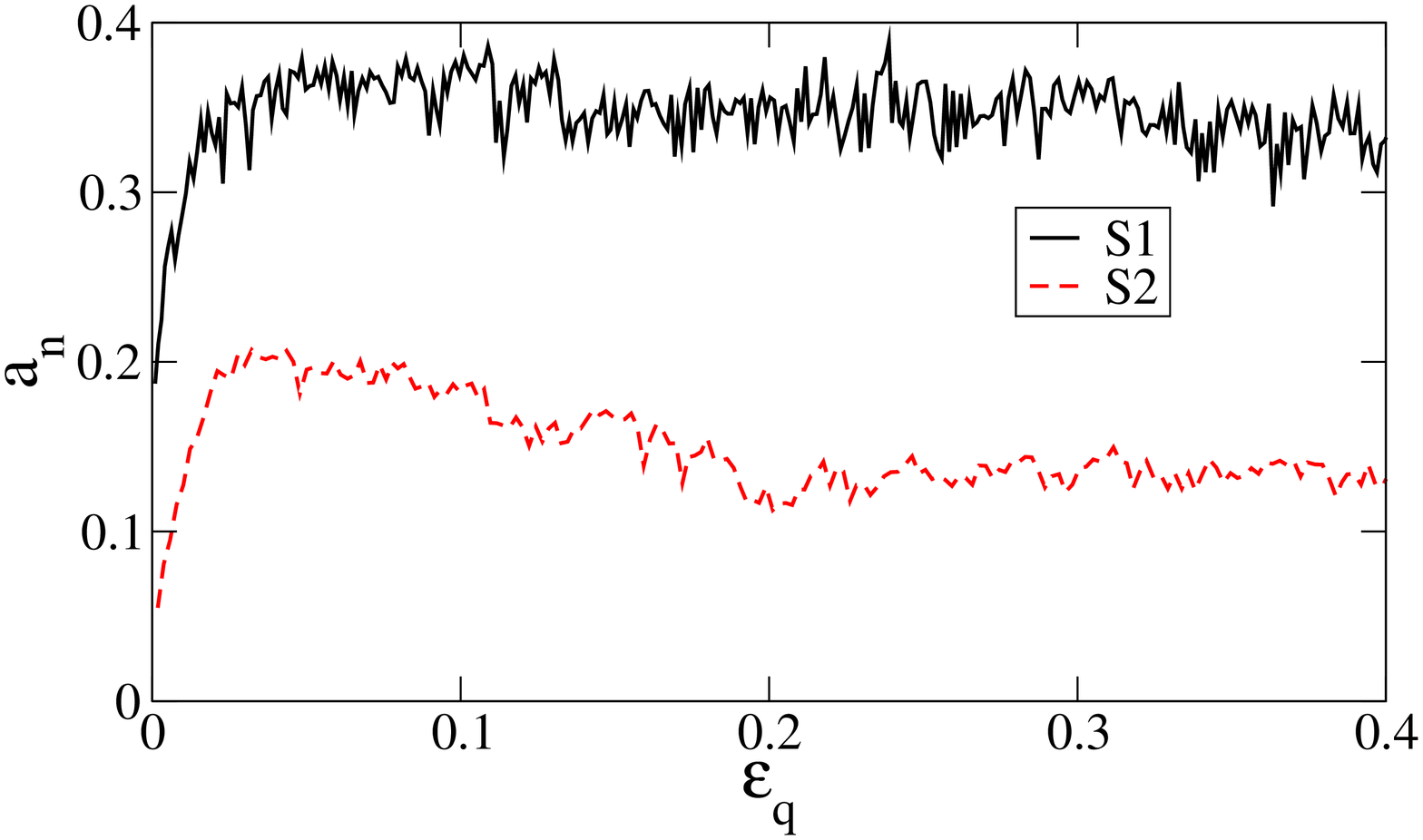}}
\centerline{\includegraphics[width=8cm]{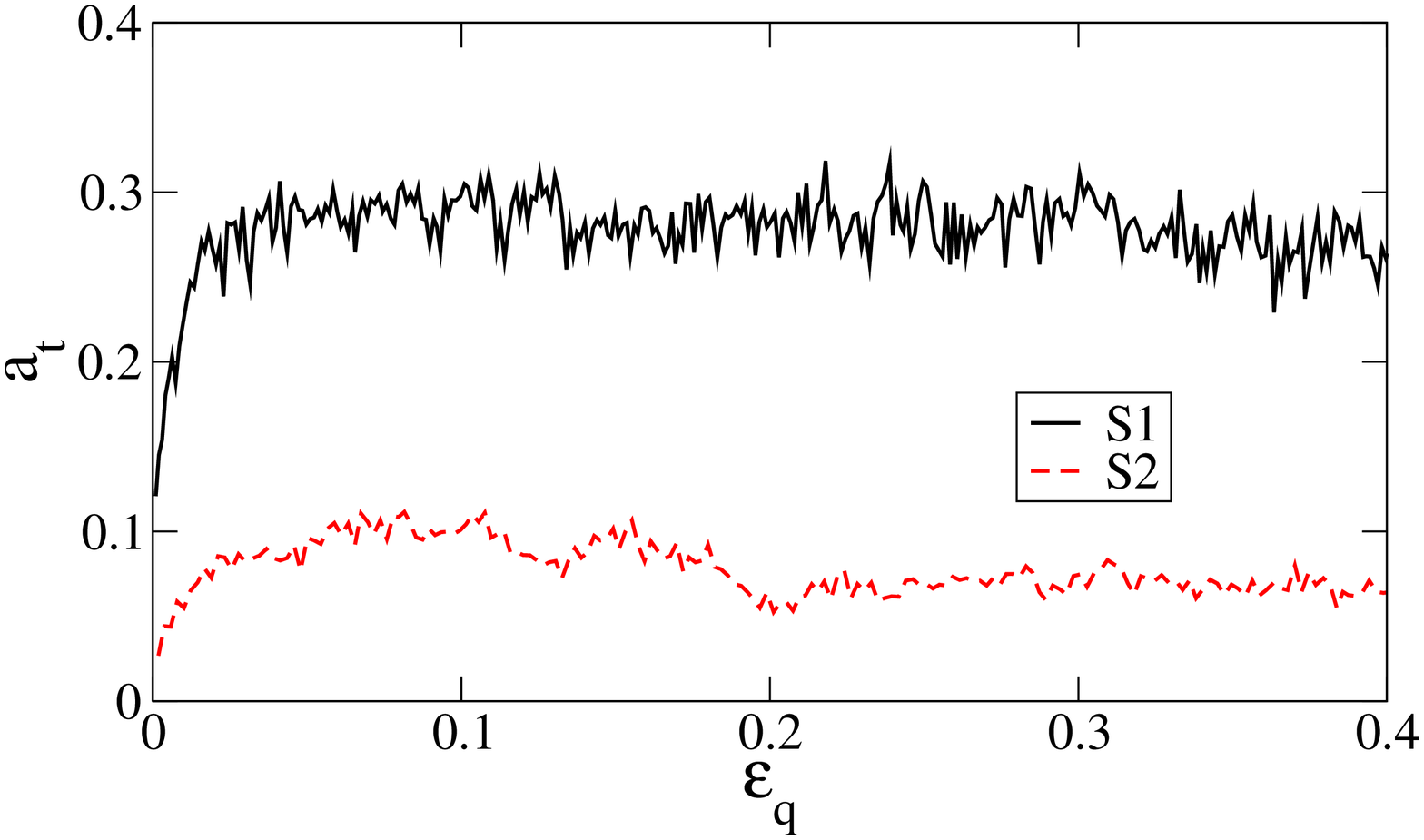}}
\caption{Evolution of force anisotropies  $a_n$ (a) and $a_t$ (b) as a function 
of cumulative shear strain  $\varepsilon_q$ in samples S1 and S2.  \label{fig:ane_ate}}
\end{figure}

Figure \ref{fig:ae} displays 
the evolution of $a$  as a function of the cumulative shear strain $\varepsilon_q$ 
in both packings. 
In both cases, $a$ increases from 0 to its largest value in the critical state.
Surprisingly, the fabric anisotropy is quite weak  in the pentagon 
packing whereas the disk packing 
is marked by a much larger anisotropy ($\simeq 0.3$).  
Fig. \ref{fig:ane_ate} shows the evolution of $a_n$ and $a_t$. 
We see that, in contrast to fabric anisotropies, 
the force anisotropies in the pentagon packing are always above those in 
the disk packing. This means that the aptitude of the pentagon packing to develop 
large force anisotropy and strong force chains is more dependent on particle 
shape than on the buildup of an anisotropic structure. 

\begin{figure}
\centerline{\includegraphics[width=8cm]{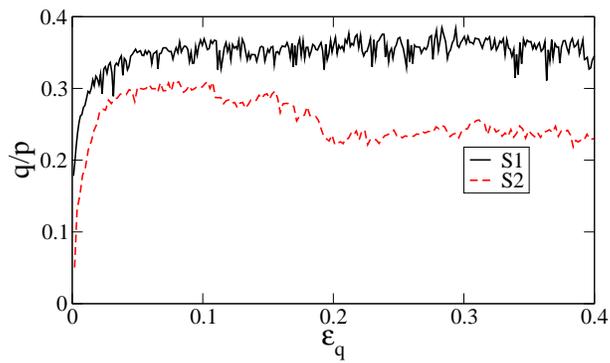}}
\caption{Normalized shear stress $q/p$ as a function of cumulative shear strain 
for the samples S1 and S2. \label{fig:qpp}}
\end{figure}

According to equation (\ref{eq:qcp}), in spite of the weak fabric anisotropy $a$, 
the larger force anisotropies $a_n$ and $a_t$ allow 
the pentagon packing  to achieve higher levels of shear strength compared to the disk packing, as shown in Fig. \ref{fig:qpp}. Our numerical data 
show that the strong force anisotropy of the polygon packing 
results from the edge-to-edge contacts that capture most strong force chains, whereas 
vertex-to-edge contacts belong mostly to the weak network. 
The pentagons  provide thus 
an interesting example where the role of fabric anisotropy 
in shear strength is marginal. Similar conclusions hold for polyhedral particles 
in 3D \cite{Azema2008}. 

The effect of the coefficient of friction $\mu_s$ between particles on the shear strength 
was discussed in section \ref{sec:friction}. The saturation of  the critical-state 
friction angle $\varphi_c$ with  increasing $\mu_s$ is related to the 
fact that, due  to disorder,  particle equilibria are fundamentally controlled 
by normal forces. Ideal situations where friction needs to be fully mobilized 
over a large number of contacts exist but are marginal. For example, a column of 
particles each with two contacts may in principal exist, but is of practically 
zero chance to occur within a disordered granular material. 
The effect of $\mu_s$ over $a_c$ manifests 
itself through $z_{min}$ which decreases with $\mu_s$. On the other hand, 
larger values of $\mu_s$ allow for  
reinforced stabilizing effect of weak contacts, increasing thereby 
force anisotropies and thus shear strength.  

The effect of adhesion is to allow for tensile forces 
mainly in the direction of extension between the particles. 
We find that the tensile forces between particles play the same 
stabilizing role with respect to the strong compressive forces as the 
weak network \cite{Radjai2001}.  
Remark that the privileged direction of weak compressive forces coincides with 
that of tensile forces. As a result, the main contribution 
to the shear strength comes from  force anisotropy. The fabric anisotropy 
is generally low and partially inhibited by the presence of adhesion.  
Note also that adhesion 
between particles involves a force scale so that its contribution  
to the shear strength is mainly expressed through  
the Coulomb cohesion $c$ (equation (\ref{eq:tau})), but it can also influence 
the internal angle of friction $\varphi_c$ through fabric anisotropy.  

The size polydispersity is an important factor that affects space-filling properties 
of granular materials. In particular, for a broad size span, the 
small particles fill and stabilize the pores between larger particles. As a result, 
larger force anisotropies and thus shear strengths are expected for 
higher levels of size polydispersity. Large particles capture strong 
force chains whereas smaller particles are mostly at the center of 
weak forces \cite{Voivret2008}. The details of force transmission and force anisotropy 
depend, however, on 
the size distribution and not only on the span. An expected effect of polydispersity 
is to allow for higher values of $z_{max}$ and thus enhanced shear strength 
as predicted by equation (\ref{eq:ac}).       

\section{Conclusion}
In this paper, we presented a brief account of 
physical mechanisms that underly the critical-state 
shear strength of granular materials. 
The short-comings of the picture of granular friction in direct analogy with 
solid friction was discussed. 
Recalling the expansion of the stress tensor 
in force and fabric anisotropies, a model was presented for 
the accessible geometrical states within a 
harmonic representation of the microstructure. 
This model, consistent with numerical simulations, 
relates the critical-state fabric anisotropy to two 
isotropic  limit states 
corresponding to the lowest and highest contact densities  
of a granular packing. The force anisotropy was analyzed 
in the light of the bimodal character of force transmission. 
It was shown that the shear strength is mainly sustained 
by the strong force network so that force anisotropy is 
mainly related to the aptitude of a granular assembly to build up 
strong force chains. Finally, the effect of material parameters 
with respect to fabric and force anisotropies was 
discussed. \\

{\em Acknowledgments.} N. Estrada and A. Taboada are acknowledged for  
figure \ref{fig:mus_muc} as well as many useful discussions about 
granular friction. I present my special thanks to S. Roux for 
interesting and inspiring ideas he has shared with me about 
the plasticity of granular media and its microscopic origins.

\bibliographystyle{plainnat}


\end{document}